\documentstyle[referee]{laa}

\begin{document}
\thesaurus{07  
	   (07.09.1; 
	    07.13.1;  
	   )}

\title{Orbital elements for motion of real particle under the action of
electromagnetic radiation}
\author{J.~Kla\v{c}ka}
\institute{Institute of Astronomy,
   Faculty for Mathematics, Physics and Informatics, \\
   Comenius University,
   Mlynsk\'{a} dolina, 842~48 Bratislava, Slovak Republic}
\date{}
\maketitle

\begin{abstract}
Discussion of different types of osculating orbital elements for motion of real
dust particle under the action of electromagnetic radiation in the central
gravitational field is presented. It is shown that physically correct
access is based on gravitational acceleration as the only radial
acceleration -- ``radiation pressure'' is not included in the radial
acceleration.


\end{abstract}

\section{Introduction}
Orbital evolution of (interplanetary) dust particle under the action
of central gravitational force and Poynting-Robertson effect
(Robertson 1937, Kla\v{c}ka 1992a) was discussed
in detail in Kla\v{c}ka (1992b -- the statement ``Equations (8) -- (11) still
hold.'' below Eq. (22) is incorrect, since Eq. (10) has to contain the product
$\mu ( 1 ~-~ \beta )$ instead of $\mu$). The paper was focused on two types
of osculating orbital elements --
the first type (``osculating elements I'') was defined
in the way that radial acceleration is given by central gravity force,
the second type (``osculating elements II'') was defined in the way that
radial acceleration is given by central gravity force together with
``radiation pressure'' force. The advantage of the second type was discussed
in Kla\v{c}ka (1992b, 1994). However, it was stressed in Kla\v{c}ka (1992b), that
osculating orbital elements of the second type ``do not take into account
correct physics''. However, mathematics has significantly simplified for
the second type of osculating orbital elements -- secular changes of the
``osculating'' orbital elements can be easily calculated.

Kla\v{c}ka (2000) has derived equation of motion for general interaction
between a particle and electromagnetic radiation. Consideration of gravitational
field of the central body (Sun in the Solar System) leads, again, to the problem
of defining of osculating orbital elements: they are defined in Kla\v{c}ka
and Kocifaj (2001 -- sections 3.2.1 and 3.2.2). Kla\v{c}ka
and Kocifaj (2001) formulate both types of osculating orbital elements and they
admit both of them as acceptable possibilities. However, some new results
presented in the form of evolution of osculating orbital elements are not
accepted by some astronomers dealing with interplanetary matter --
they insist on using of the second type of osculating elements, i. e., the
use of the osculating orbital elements II (section 3.2.2 in Kla\v{c}ka and
Kocifaj 2001) is the only correct method according to their opinion.

The aim of this contribution is to look at the problem which type
of osculating orbital elements is correct -- if the ``radial radiation pressure''
has to be considered together with the central gravitational force or not.

\section{Osculating orbital elements -- definition}
As a definition of osculating orbital elements we take the definition presented
in Brouwer and Clemence (1961), since this book is very often used by
scientists dealing with celestial mechanics and its application to Solar
System. Brouwer and Clemence write (p. 273): ``As the motion progresses
under the influence of the various attracting bodies, the coordinates and
velocity components at any instant may be used to obtain a set of six
orbital elements. These are precisely the elements of the ellipse that the
body would follow if from that particular instant on, the accelerations
caused by all ``perturbing'' bodies ceased to exist.'' As for our purposes,
it is sufficient to make a small change: ``attracting bodies'' are replaced
by ``forces''. Our force coresponds to the electromagnetic radiation force.

\section{Osculating orbital elements for particle interacting with
electromagnetic radiation}
Equation of motion for a particle in the central gravitational and
electromagnetic radiation fields may be written (to the first order in
$\vec{v}/c$ -- higher orders are neglected, where $\vec{v}$ is velocity
of the particle, $c$ is the speed of light) as
\begin{eqnarray}\label{1}
\frac{d~ \vec{v}}{d ~t} &=&  -~\frac{\mu}{r^{3}} ~ \vec{r} ~+~
		 \frac{S ~A'}{m~c} ~ \left \{ Q_{R} ' ~ \left [
		 \left ( 1~-~ \vec{v} \cdot \hat{\vec{S}_{i}} / c \right ) ~
		 \hat{\vec{S}_{i}} ~-~ \vec{v} / c \right ] ~+~ \right .
\nonumber \\
& &  \left .  \sum_{j=1}^{2} ~Q_{j} ' ~\left [  \left ( 1~-~ 2~
	      \vec{v} \cdot \hat{\vec{S}_{i}} / c ~+~
	      \vec{v} \cdot \hat{\vec{e}_{j}} / c \right ) ~ \hat{\vec{e}_{j}}
	      ~-~ \vec{v} / c \right ] \right \} ~,
\end{eqnarray}
where $\mu = G~ ( M~+~m )$, $G$ is gravitational constant,
$M$ is mass of the central body, $m$ is mass of the particle, $S$ is
flux density of radiation energy, $Q'_{R}$, $Q'_{1}$, $Q'_{2}$ are
``effective factors'', $A'$ is geometrical cross-section of a sphere
of volume equal to the volume of the particle, $\vec{r}$ is position vector
of the particle with respect to the body of mass $M$,
$\hat{\vec{S}_{i}} \equiv \vec{r} / | \vec{r} |$ -- for more details
see Kla\v{c}ka (2000) or Kla\v{c}ka and Kocifaj (2001).

Eq. (1) may be rewritten in the form
\begin{eqnarray}\label{2}
\frac{d~ \vec{v}}{d ~t} &=&  -~\frac{\mu \left ( 1 ~-~ \beta \right )}{r^{3}}
			       ~ \vec{r} ~+~  \vec{a}_{D} ~,
\nonumber \\
\vec{a}_{D} &\equiv& \frac{S ~A'}{m~c} ~ \left \{ -~ Q_{R} ' ~ \left [
		 \left ( \vec{v} \cdot \hat{\vec{S}_{i}} / c \right ) ~
		 \hat{\vec{S}_{i}} ~+~ \vec{v} / c \right ] ~+~ \right .
\nonumber \\
& &  \left .  \sum_{j=1}^{2} ~Q_{j} ' ~\left [  \left ( 1~-~ 2~
	      \vec{v} \cdot \hat{\vec{S}_{i}} / c ~+~
	      \vec{v} \cdot \hat{\vec{e}_{j}} / c \right ) ~ \hat{\vec{e}_{j}}
	      ~-~ \vec{v} / c \right ] \right \} ~,
\nonumber \\
\beta &\equiv& Q_{R} ' ~ \frac{S ~A'}{m~c} ~/~ \frac{\mu}{r^{2}} ~.
\end{eqnarray}

General opinion is that osculating orbital elements have to be taken with
respect to the total central acceleration -- given by the factor
$\mu ( 1 ~-~ \beta )$. This corresponds to disturbing acceleration
$\vec{a}_{D}$ in Eq. (2). However, this access is not correct. We want to show
this. Let the disturbing acceleration $\vec{a}_{D}$ ceases at any instant.
The problem is, that solution of Eq. (2) for $\vec{a}_{D} =$ 0 does not
correspond to ellipse (or to any kind of conic section) -- $\beta -$ parameter
is not a constant and its value changes during the motion even for the case
$\vec{a}_{D} =$ 0. As a consequence, the only correct form of defining
osculating orbital elements is defined by central acceleration given by the
factor $\mu$.

\section{Discussion}
If we want to use osculating orbital elements in the way as they are
defined in celestial mechanics, we have to use the central unperturbed
acceleration given by gravitational acceleration of the central body --
radial component of the radiation pressure cannot be included. Since
the values of orbital elements exhibit a large dispersion
during a revolution around the central body (Kla\v{c}ka 1994),
it is wise -- e. g., for the purpose of figures -- to make time average
(say, from pericenter to the following pericenter).

Of course, it is possible to use the central acceleration
containing also radial radiation pressure. However, in this case, it is
not sufficient to use only `orbital' elements for finding $\vec{r}$ and $\vec{v}$
-- we need also to know $\beta$ for a given instant; `orbital' elements
have no sense of osculating orbital elements defined in celestial mechanics.
Time averaging is also recommended.

\section{Conclusion}
The problem of osculating orbital elements for the case of motion of a particle
under the action of electromagnetic radiation was presented.
We have shown that dealing with osculating orbital elements as defined
in celestial mechanics, there is only one correct definition, in general
case: central acceleration is given by gravitational acceleration, radial
radiation pressure is not included.

\acknowledgements{}
The paper was partially supported by the Scientific Grant Agency VEGA,
grant No. 1/7067/20.

\end{document}